\documentclass[%
 amsmath,amssymb,
reprint,%
]{revtex4-1}

\usepackage{graphicx}% Include figure files
\usepackage{dcolumn}% Align table columns on decimal point
\usepackage{bm}% bold math
\usepackage{color}
\begin{document}

\title[Sample title]{Quantum Hall resistance standards from graphene grown by\\ chemical vapour deposition on silicon carbide}% Force line breaks with \\
%\thanks{Footnote to title of article.}

\author{F. Lafont$^1$, R. Ribeiro-Palau$^1$, D. Kazazis$^2$,  A. Michon$^3$,  O. Couturaud$^4$, C. Consejo$^4$, T. Chassagne$^5$, M. Zielinski$^5$,  M. Portail$^3$, B. Jouault$^4$, F. Schopfer$^1$ and W. Poirier$^1$}
\email{wilfrid.poirier@lne.fr.}

\affiliation{$^1$LNE - Laboratoire National de M\'{e}trologie et d'Essais, Quantum electrical metrology department, avenue Roger Hennequin, 78197 Trappes, France}
\affiliation{$^2$LPN - Laboratoire de Photonique et de Nanostructures,CNRS, Route de Nozay, 91460  Marcoussis, France}
\affiliation{$^3$CRHEA - Centre de Recherche sur l'H\'et\'ero\'epitaxie et ses Applications, CNRS, rue Bernard Gr\'egory, 06560 Valbonne, France}
\affiliation{$^4$L2C - Laboratoire Charles Coulomb, UMR 5221 CNRS-Université de Montpellier, Place Eug\`ene Bataillon, 34095 Montpellier, France}
\affiliation{$^5$NOVASiC, Savoie Technolac, Arche Bat 4, 73375 Le Bourget du Lac, France}
\date{\today}% It is always \today, today,

%Use showkeys class option if keyword

\begin{abstract}
\textbf{Replacing GaAs by graphene to realize more practical quantum Hall resistance standards (QHRS), accurate to within $10^{-9}$ in relative value, but operating at lower magnetic fields than 10 T, is an ongoing goal in metrology. To date, the required accuracy has been reported, only few times, in graphene grown on SiC by sublimation of Si, under higher magnetic fields. Here, we report on a device made of graphene grown by chemical vapour deposition on SiC which demonstrates such accuracies of the Hall resistance from 10 T up to 19 T at 1.4 K. This is explained by a quantum Hall effect with low dissipation, resulting from strongly localized bulk states at the magnetic length scale, over a wide magnetic field range. Our results show that graphene-based QHRS can replace their GaAs counterparts by operating in as-convenient cryomagnetic conditions, but over an extended magnetic field range. They rely on a promising hybrid and scalable growth method and a fabrication process achieving low-electron density devices.}%\newpage
\end{abstract}

\maketitle

The metrology of the resistance unit has been continuously progressing since the discovery that the transverse resistance of a two-dimensional electron gas (2DEG) in a perpendicular magnetic field is quantized at universal values $R_\mathrm{K}/i$, where $R_\mathrm{K}\equiv h/e^2$ is the von Klitzing constant, $h$ the Planck constant, $e$ the electron charge, and $i$ an integer\cite{Klitzing1980}. Using GaAs-based heterostructures to form the 2DEG, it has been possible to develop QHRS reproducing $R_\mathrm{K}/2$ with a relative uncertainty down to $3\times10^{-11}$\cite{Schopfer2013}, as well as accurate low and high resistance QHRS based on arrays of Hall bars\cite{Poirier2004} and QHRS adapted to the alternating current (AC) regime\cite{AhlersAC2009}.  More recently, it has been considered by metrologists that an accurate QHRS could be developed in graphene\cite{Poirier2009,Schopfer2012}, possibly surpassing the usual GaAs based ones.

The Dirac physics in monolayer graphene manifests itself by a quantum Hall effect (QHE)\cite{Novoselov2005,Zhang2005} with Landau levels (LLs) at energies $\pm v_{\mathrm{F}}\sqrt{2\hbar n e B}$ with a $4eB/h$ degeneracy (valley and spin) and a sequence of Hall resistance plateaus at $R_{\mathrm{H}}=\pm R_{\mathrm{K}}/(4(n+1/2))$ (with $n\geqslant 0$)\cite{Gusynin2005}. The energy spacing between the two first LLs, $\Delta E(B)= 36\sqrt{B[\mathrm{T}]}~\mathrm{meV}$, is much larger than in GaAs ($1.7B[\mathrm{T}]~\mathrm{meV}$) for currently accessible  magnetic fields. It results that the $\nu=2$ Hall resistance plateau of value $R_{\mathrm{K}}/2$ ($\nu=hn_\mathrm{s}/eB$ is the Landau level filling factor and $n_\mathrm{s}$ the carrier density) can be observable even at room temperature\cite{Novoselov2007}. This opens the way towards a more convenient QHRS in graphene\cite{Poirier2009,Poirier2010} operating at lower magnetic fields ($B\leq$ 4 T), higher temperatures ($T\geq$ 4 K) and higher measurement currents ($I\geq100~\mu$A) compared to its GaAs counterpart. From previous measurements of the Hall resistance quantization in graphene produced by various methods (exfoliation of graphite, chemical vapour deposition (CVD) on metal, and Si sublimation from SiC), it was concluded that the production of graphene-based QHRS (G-QHRS) requires large graphene monolayers (few $\mathrm{10000~\mu m^{2}}$) with homogeneous low carrier densities ($<2\times 10^{11}\mathrm{cm^{-2}}$) and high carrier mobilities ($\geq5000~\mathrm{cm^{2}V^{-1}s^{-1}}$). Thus, although the quantized Hall resistance was measured with a relative standard uncertainty of $6.3\times10^{-9}$ (1 s.d.), on the $\nu=2$ plateau at \emph{B}=18 T and \emph{T}=60 mK, in monolayer graphene obtained by mechanical exfoliation\cite{Wosczczyna2012}, this technique was quickly discarded because it produces few monolayers of rather small size and it lacks reproducibility\cite{Giesberg2009,Wosczczyna2012,Guignard2012,Schopfer2012}. In the case of graphene grown by CVD on metal and transferred on a $\mathrm{SiO_2/Si}$ substrate, the Hall resistance was measured far from being correctly quantized\cite{Shen2011,Lafont2014}. It was shown that the presence of grain boundaries and wrinkles jeopardized the quantization\cite{Lafont2014,Cummings2014}. As far as we know, no accurate measurement of $R_{\mathrm{H}}$ has been reported in graphene grown by CVD on metal. The quantized Hall resistance was measured by metrologists of the National Physical Laboratory in the United Kingdom (NPL), on the $\nu=2$ plateau, in monolayer graphene grown by Si sublimation on the silicon face of SiC, produced by Link\"oping University. The agreement of the Hall resistance with $R_{\mathrm{K}}/2$ was demonstrated with a relative standard measurement uncertainty, down to $8.7\times10^{-11}$ at \emph{B}=14 T and \emph{T}=0.3 K\cite{Tzalenchuk2010,Janssen2011}, and slightly lower than $10^{-9}$ at \emph{B}=11.5 T and \emph{T}=1.5 K\cite{Janssen2012}. The NPL work also showed the flatness of the quantized Hall resistance plateau with a relative uncertainty of a few $10^{-9}$, over a 2.5 T range extending from 11.5 T to 14 T\cite{Janssen2012}. In lower carrier-density samples produced by Graphensic AB, a spin-off from Link\"oping University research, the Hall resistance was measured on the $\nu=2$ plateau at lower magnetic fields in the range from 2 T to 8 T. However, the accuracy of the quantized Hall resistance was not demonstrated with a relative uncertainty better than a few ${10^{-7}}$ at 3 T and $\approx1.8\times{10^{-8}}$ at 8 T\cite{Satrapinski2013}. Moreover, a large dispersion of the measurements (up to $0.5\times{10^{-6}}$ in relative value) was observed by changing the Hall terminal pairs used, manifesting strong inhomogeneities in the samples. It turns out that no other $10^{-9}$-accurate QHRS (a QHRS accurate to within $10^{-9}$ in relative value) was achieved from any other graphene sources, though large scale high mobility graphene was produced\cite{Pallecchi2014}. Thus, only a few samples from a unique material supplier demonstrated the accuracy required in national metrology institutes, which compromises the sustainability of G-QHRS. Moreover, it was obtained under experimental conditions less convenient than those of currently used GaAs-based QHRS (GaAs-QHRS).
\begin{figure}[h!]
\begin{center}
\includegraphics[width=3.3in]{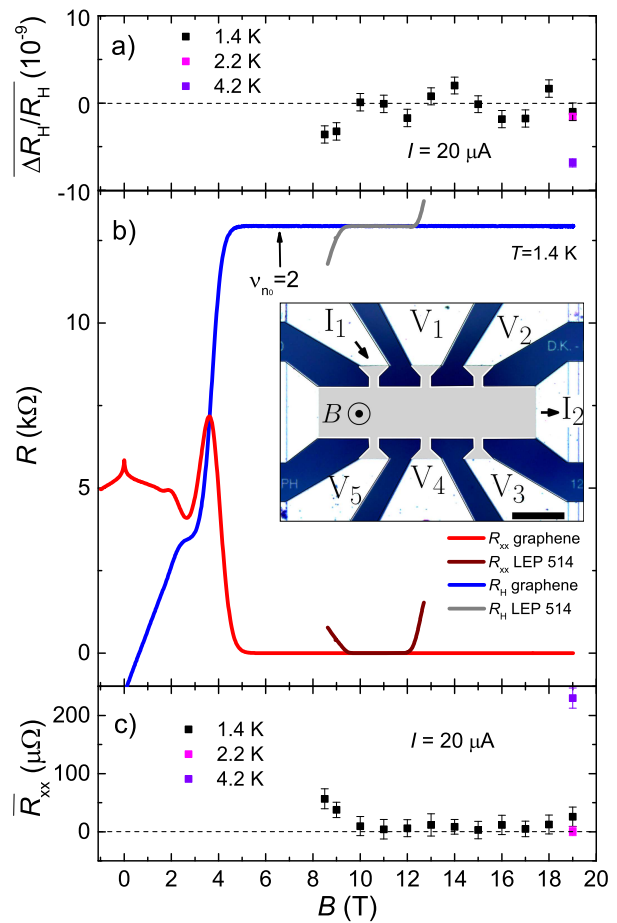}
\caption{\textbf{Magnetic field dependence of the Hall quantization in graphene grown by CVD on SiC}. a) Hall resistance deviation $\overline{\Delta R_\mathrm{H}/R_\mathrm{H}}$ measured on the $\nu=2$ plateau at 1.4 K (black), 2.2 K (magenta) and 4.2 K (violet). b) Longitudinal ($R_\mathrm{xx}$) and Hall ($R_\mathrm{H}$) resistances (a $\mathrm{100~nA}$ current $I$ circulates between $\mathrm{I_1}$ and $\mathrm{I_2}$ terminals, the longitudinal and Hall voltages are measured using ($\mathrm{V_1}$, $\mathrm{V_2}$) and ($\mathrm{V_2}$, $\mathrm{V_3}$) terminal pairs, respectively) for \emph{B} varying from -1 T to 19 T for the graphene sample (red and blue curves, respectively) and varying from 8 T to 13 T for the GaAs sample (wine and grey curves respectively). $\nu_{n_0}$ is the Landau level filling factor calculated from the carrier density $n_0$ determined at low magnetic fields. Inset: optical image of the sample with terminal labels, scale bar, 100 $\mathrm{\mu m}$. A very wide Hall resistance plateau is observed in the large Hall bar device. c) Accurate measurements of the longitudinal resistance $\overline{R_\mathrm{xx}}$ \emph{versus} \emph{B} at 1.4 K (black), 2.2 K (magenta) and 4.2 K (violet). Error bars represent combined standard uncertainties, given with a coverage factor k=1 corresponding to one standard deviation (1 s.d.). A perfect quantization of the Hall resistance, without significant deviations with regards to the relative standard measurement uncertainty of $10^{-9}$, is observed over a magnetic field range of 9 T from 10 T. It coincides with $\overline{R_\mathrm{xx}}$ values lower than $\mathrm{(30\pm20)~\mu \Omega}$. }\label{fig1}
\end{center}
\end{figure}

Here, we report on a G-QHRS made of monolayer graphene grown by propane/hydrogen CVD on SiC\cite{Michon2010}, a hybrid technique which allows the tuning of the electronic transport properties \cite{Jabakhanji2014}. The Hall resistance, measured on the $\nu=2$ plateau with a $10^{-9}$ relative standard measurement uncertainty, is found in agreement with $R_{\mathrm{K}}/2$ over a 9 T-wide magnetic field range from \emph{B}=10 T up to \emph{B}=19 T at \emph{T}=1.4 K. These cryomagnetic experimental conditions overlap, for the first time, that of the reference GaAs-QHRS used, making this device an operational QHRS substitute in current set-ups used in national metrology institutes. The relative discrepancy between the quantized Hall resistance of the G-QHRS and the GaAs-QHRS is found equal to $(-2\pm 4)\times 10^{-10}$, which constitutes a new proof of the universality of the QHE. The QHE physics of the large $\nu=2$ Hall resistance plateau is investigated using accurate measurement techniques based on specialized metrological instruments. It turns out that the dissipation is dominated by the variable range hopping (VRH) mechanism. The wide quantized Hall resistance plateau is characterized by a localization length of states at Fermi energy that remains very close to the magnetic length over a large magnetic field range of 9 T. This is likely caused by a pinning of the LL filling factor at $\nu=2$ due to a charge transfer from the donor states in the interface layer between SiC and graphene. The measurement of a second $10^{-9}$-accurate G-QHRS fabricated from a different growth run, in similar cryomagnetic conditions as the GaAs-QHRS establishes a worthy repeatability of the propane/hydrogen CVD on SiC growth method. Initiated in 2010\cite{Michon2010}, this production technique is now mature and very promising to develop a challenging G-QHRS surpassing the GaAs-QHRS in the near future.

\subsection*{Results}
\subparagraph*{Magneto-resistance characterizations.}

 Figure \ref{fig1}-b) shows the Hall resistance $R_\mathrm{H}$ and the longitudinal resistance per square $R_\mathrm{xx}$ measured as a function of the magnetic field \emph{B} at a temperature \emph{T}=1.4 K and a measurement current \emph{I} of 100 nA in a large 100 $\mu$m by 420 $\mu$m Hall bar sample, insert Fig. \ref{fig1}-b), made of graphene grown on the Si face of SiC by propane/hydrogen CVD under a mixture of propane, hydrogen and argon\cite{Michon2010,Michon2013} (see Methods). At low magnetic fields, from the Hall slope and the Drude resistivity, a low electron density $n_0=\mathrm{3.2\times10^{11}cm^{-2}}$ and an electronic mobility $\mu=\mathrm{\mathrm{3500~cm^2V^{-1}s^{-1}}}$ are calculated. At higher magnetic fields, a wide Hall resistance plateau $R_\mathrm{K}/2$ can be observed, from \emph{B}=5 T up to \emph{B}=19 T (maximum accessible magnetic field in our set-up) and coinciding with a dropping to zero of $R_\mathrm{xx}$. It extends far beyond the magnetic field \emph{B}= 6.6 T at which $\nu_{n_0}=2$, where $\nu_{n_0}$ is the LL filling factor calculated from the carrier density determined at low magnetic fields. One can also notice the $R_\mathrm{K}/6$ Hall resistance plateau between 2 T and 3 T. The comparison is striking when one compares the $R_\mathrm{K}/2$ Hall resistance plateau with the one of the most widespread in national metrology institutes GaAs-based QHR (LEP514\cite{Piquemal1993}) which only extends over 2 T starting from 10 T (green curve). Such robust QHE, characterized by wide $\nu =2$ Hall resistance plateau observable from low magnetic fields, were reproduced in other Hall bar samples fabricated from graphene grown by CVD on SiC, as discussed in Methods.

\begin{figure}[h!]
\begin{center}
\includegraphics[width=3.3in]{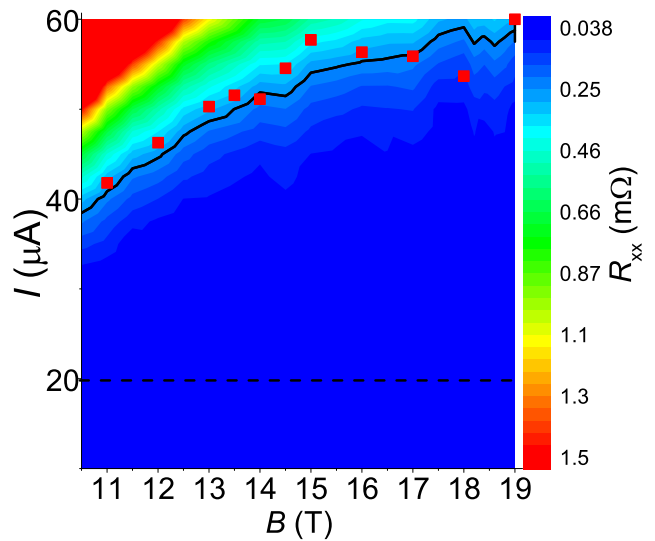}
\caption{\textbf{Dissipation in (\emph{B},\emph{I}) space in the $\nu=2$ plateau}. Color rendering of the longitudinal resistance per square $R_\mathrm{xx}$ (measured using terminal-pair ($\mathrm{V_3}$,$\mathrm{V_5}$)) as a function of \emph{I} (circulating between $\mathrm{I_1}$ and $\mathrm{I_2}$ terminals) and \emph{B} at $T=1.4$ K. $I_C(B)$ (black solid line) corresponds to the evolution of the critical current (breakdown current) leading to $R_\mathrm{xx}=0.25~\mathrm{m}\Omega$ as a function of \emph{B}. The horizontal black dashed line indicates the current used for the accurate measurements. $I_\mathrm{C}(B)$ can be well adjusted by $\xi(B)^{-2}$ data (red square), where $\xi(B)$ is the localization length. There is no significant dissipation in the QHE regime ($R_\mathrm{xx}<0.25~\mathrm{m}\Omega$) for currents lower than 40 $\mu$A in the magnetic field range from 10.5 T to 19 T.}\label{fig2}
\end{center}
\end{figure}

Figure \ref{fig2} shows the color rendering of the longitudinal resistance per square $R_\mathrm{xx}$ measured as a function of the measurement current $I$ and the magnetic field $B$. It shows that the 2D electron gas is not significantly dissipative ($R_\mathrm{xx}<0.25~\mathrm{m}\Omega$) for currents as high as $40~\mu$A in the large range of magnetic fields between 10.5 T and 19 T. We also carried out three-terminal measurements of contact resistances in the $R_\mathrm{K}/2$ Hall plateau. All contacts between metallic pads and graphene, except one (left terminal of the Hall bar, inset Fig. \ref{fig1}-b)), are ohmic with a resistance lower than $1~\Omega$. That's why in all the reported four-probes resistance measurements, the current $I$ circulates between I$_1$ and I$_2$ terminals. For a current of $20~\mu$A, lower than the aforementioned limit of $40~\mu$A, we performed accurate four-probes measurements of $R_\mathrm{H}$ and $R_\mathrm{xx}$ (see Methods). Two Hall resistances are determined using Hall terminal-pairs ($\mathrm{V_1}$,$\mathrm{V_4}$) and ($\mathrm{V_2}$,$\mathrm{V_3}$).  Two longitudinal resistances are determined using longitudinal terminal-pairs, located on both edges of the Hall bar, ($\mathrm{V_1}$,$\mathrm{V_2}$) and ($\mathrm{V_3}$,$\mathrm{V_5}$). Longitudinal resistances are normalized to a square.

\subparagraph*{Resistance quantization.}

$R_\mathrm{H}$ is indirectly compared to the $R_\mathrm{K}/2$ value given by a reference GaAs-QHRS (LEP514) using a $100~\Omega$ transfer resistor. The comparison is performed using a resistance bridge equipped with a cryogenic current comparator (CCC). To minimize resistance comparison errors , the same measurement current and settings of the bridge are used to calibrate the 100-$\Omega$ resistor either from the G-QHRS or the GaAs-QHRS. All uncertainties reported in the following are expressed as one standard deviation (1 s.d.). Let us note $\Delta R_\mathrm{H}/R_\mathrm{H}$ the relative deviation of the Hall resistance $R_\mathrm{H}$ from $R_\mathrm{K}/2$ ($\Delta R_\mathrm{H}=R_\mathrm{H}-R_\mathrm{K}/2$). $\overline{\Delta R_\mathrm{H}/R_\mathrm{H}}$ and $\overline{R_{xx}}$ are obtained from the mean value of the measurements performed using the two Hall terminal-pairs and the two longitudinal terminal-pairs previously mentioned. Figure \ref{fig1}-a) reports $\overline{\Delta R_\mathrm{H}/R_\mathrm{H}}$ values determined with a combined standard measurement uncertainty close to $1\times10^{-9}$ (the main contribution comes from the instability of the $100~\Omega$ transfer resistor) as a function of \emph{B} for a measurement current of $\mathrm{20~\mu A}$ and a temperature \emph{T}=1.4 K. It shows a perfect quantization of the Hall resistance with no significant deviations over the whole magnetic field range of 9 T between \emph{B}=10~T and \emph{B}=19~T, which coincides with $\overline{R_\mathrm{xx}}$ values lower than $\mathrm{(30\pm20)~\mu \Omega}$ (see Fig. \ref{fig1}-c)): $\overline{\Delta R_\mathrm{H}/R_\mathrm{H}}$ discrepancies are all within the expanded (k=2) standard measurement uncertainties (2 s.d.), where a coverage factor k=2 gives an expected confidence level of $95~\%$. More sensitive measurements performed with the CCC (see Methods) show that $\overline{R_\mathrm{xx}}$ amounts to $(10.5\pm 2.4)~\mu\Omega$ and $(1.2\pm 1.7)~\mu\Omega$ at \emph{B}=10 T and \emph{B}=19 T respectively, demonstrating a very low dissipation level in the graphene electron gas. Measurements show that the $R_\mathrm{H}$ values determined from the two Hall terminal-pairs are in agreement within a relative measurement uncertainty close to $\approx 1\times10^{-9}$, demonstrating the homogeneity of the Hall quantization in the sample over a large surface. The mean value of $\overline{\Delta R_\mathrm{H}/R_\mathrm{H}}$ measurements carried out at magnetic fields between 10 T and 19 T is $-2\times 10^{-10}$ covered by an experimental standard deviation of the mean of $4\times 10^{-10}$. Below \emph{B}=10~T, the Hall resistance starts to deviate from the quantized value and the longitudinal resistance significantly increases. Fig. \ref{fig1}-a) also reports that $\overline{\Delta R_\mathrm{H}/R_\mathrm{H}}$ is equal, at \emph{B}=19 T, to $(-1.5\pm0.4)\times 10^{-9}$ and $(-7\pm0.5)\times 10^{-9}$ at \emph{T}=2.2 K and \emph{T}=4.2 K respectively. This is due to an increase of $\overline{R_\mathrm{xx}}$ reaching $\mathrm{0.23~m\Omega}$ at \emph{T}=4.2 K (see Fig. \ref{fig1}-c)). These Hall quantization measurements show first of all that the G-QHRS can operate accurately at magnetic fields as low as the ones of the reference GaAs-QHRS ($10~\mathrm{T}\leq\ B \leq 11~\mathrm{T}$) with a similar temperature of 1.4 K. This demonstrates that this G-QHRS can directly replace a GaAs-QHRS in a conventional QHE setup of a national metrology institute equipped with a 12 T magnet. The repeatability of the CVD on SiC growth method and of the technological process to obtain such competitive G-QHRS was tested by the measurement of another Hall bar device, having the same geometry, made of graphene produced in a different growth run (several months later). At $T=1.3$ K, the Hall resistance measured in this second G-QHRS device is also in agreement with $R_\mathrm{K}/2$ at \emph{B}=10 T within a relative measurement uncertainty below $10^{-9}$ since $\Delta R_\mathrm{H}/R_\mathrm{H}=(4\pm8)\times10^{-10}$. As in the main G-QHRS, the flatness of the $\nu=2$ Hall resistance plateau was demonstrated above 10 T over a magnetic field range larger than in GaAs-based devices: at \emph{B}=10.8 T and at \emph{B}=12 T, $\Delta R_\mathrm{H}/R_\mathrm{H}$ is found equal to $(7\pm8)\times10^{-10}$ and $(-2\pm8)\times10^{-10}$, respectively. This demonstrates a notable degree of reproducibility of the QHRS fabrication process which is an asset for the resistance metrology application.

\begin{figure}[h!]
\begin{center}
\includegraphics[width=3.3in]{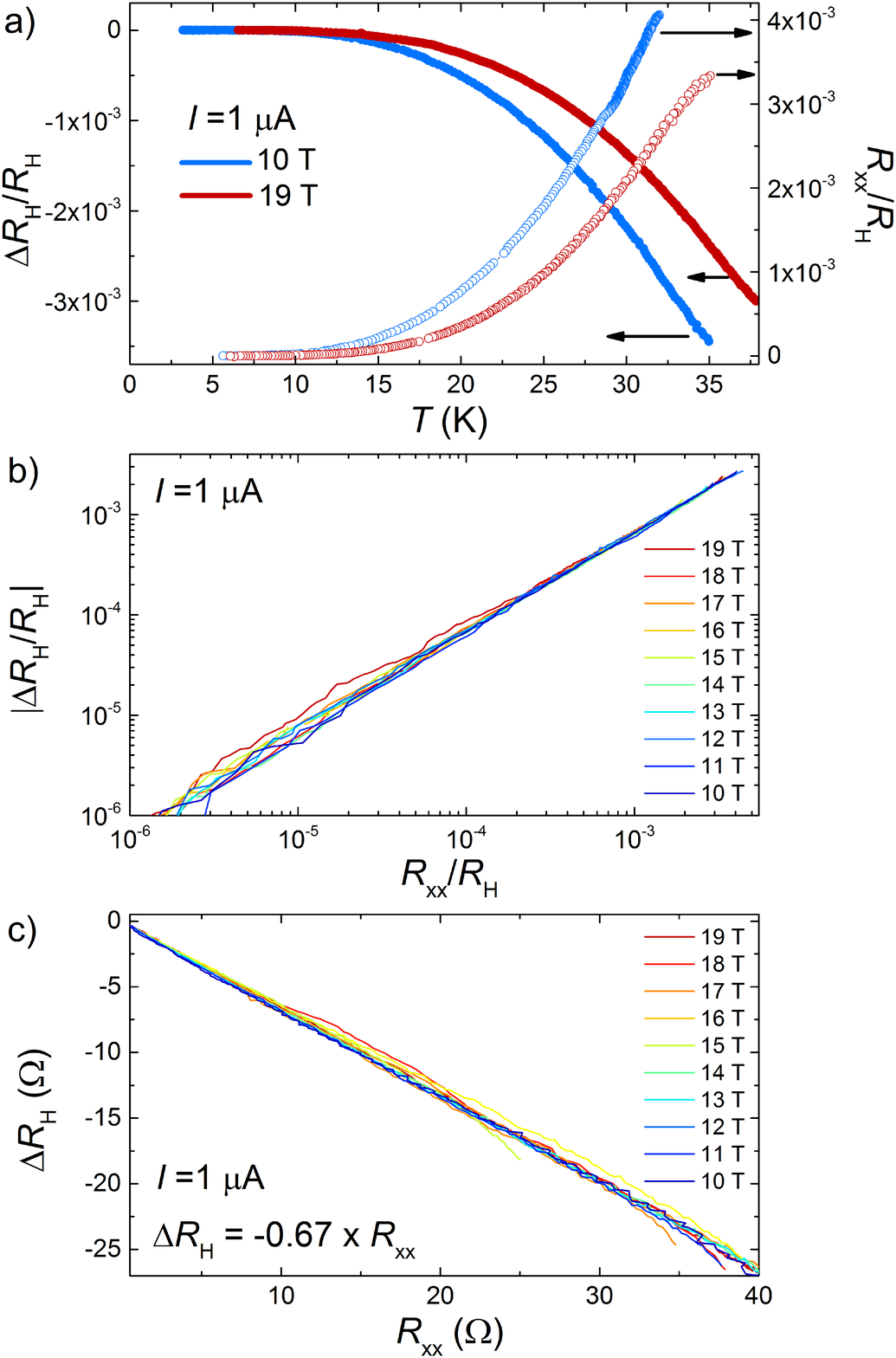}
\caption{\textbf{Relationship between Hall and longitudinal resistances}. a) $\Delta R_\mathrm{H}/R_\mathrm{H}$ (filled circles, left axis) and $R_\mathrm{xx}/R_\mathrm{H}$ (empty circles, right axis) as a function of the temperature $T$ for two magnetic fields \emph{B}=10 T (blue) and \emph{B}=19 T (red). b) $|\Delta R_\mathrm{H}/R_\mathrm{H}|$ as a function of $R_\mathrm{xx}/R_\mathrm{H}$ for several magnetic fields \emph{B} in log-log scale. c) $\Delta R_\mathrm{H}$ as a function of $R_\mathrm{xx}$ in linear scales. A linear relationship, $\Delta R_{\rm H}= -0.67\times R_{\rm xx}$ independent of the magnetic field in the range from 10 T to 19 T, is found. It is concluded that relative deviations of the Hall resistance from $R_\mathrm{K}/2$ are smaller than $10^{-9}$ for longitudinal resistance values lower than 15 $\mu\Omega$.}\label{fig3}
\end{center}
\end{figure}

In the first presented G-QHRS, the relationship between $R_\mathrm{H}(T)$ and $R_\mathrm{xx}(T)$ was investigated in the range from 4 K to 40 K by performing measurements (using one Hall terminal-pair ($\mathrm{V_1}$,$\mathrm{V_4}$) and one longitudinal terminal-pair ($\mathrm{V_3}$,$\mathrm{V_5}$)) at magnetic fields between \emph{B}=10 T and \emph{B}=19 T using a low AC (2~Hz frequency) current of $1~\mu$A. From measurements of both $R_\mathrm{H}(T)$ and $R_\mathrm{xx}(T)$ (Fig. \ref{fig3}-a)), one can report $\Delta R_\mathrm{H}/R_\mathrm{H}(T)$ as a function of $R_\mathrm{xx}(T)/R_\mathrm{H}(T)$ for different \emph{B} values in log-log scale (see Fig. \ref{fig3}-b)). The curves carried out at different \emph{B} values are superimposed on a single straight line of unitary slope over four decades of $R_\mathrm{xx}$. This corresponds to a relationship $\Delta R_\mathrm{H}=-0.67\times R_\mathrm{xx}$ (see Fig. \ref{fig3}-c)). The same relationship is found by varying the magnetic field at a given temperature (not shown). We can therefore conclude that relative deviations of the Hall resistance from $R_\mathrm{K}/2$ are smaller than $10^{-9}$ for $R_\mathrm{xx}$ values lower than $15~\mu\Omega$. The longitudinal resistance values of $(10.5\pm 2.4)~\mu\Omega$ and $(1.2\pm 1.7)~\mu\Omega$ measured at \emph{B}=10 T and \emph{B}=19 T respectively should lead to small relative discrepancies to $R_\mathrm{K}/2$ of $\approx 6\times10^{-10}$ and less than $\approx 1\times10^{-10}$ respectively, thus experimentally not observable in Fig.\ref{fig1}. The linear relationship between $R_\mathrm{H}$ and $R_\mathrm{xx}$ can be described by an effective geometric coupling. In GaAs-QHRS, this coupling is usually explained by the finite width of the voltage terminal arms with respect to the Hall bar channel\cite{VanderwelPhd1988,Poirier2009} or the inhomogeneous circulation of the current (for example due to the residual inhomogeneity of the carrier density)\cite{DominguezPhd1988, Vanderwel1988}. The first mechanism would lead to a coupling factor of $(-l/W)=-0.2$, where $l=\mathrm{20~\mu m}$ is the width of the voltage arm and $W=\mathrm{100~\mu m}$ is the width of the Hall bar channel. On the other hand, the specific injection of the current by the I$_1$ terminal could explain the larger observed coupling. Another explanation relies on the impact of SiC steps oriented at $45^{\circ}$ with respect to the Hall bar orientation, with the presence of bilayer patches along them, that can cause a tilted circulation of the current. Representing about $10~\%$ of the total surface in this sample, these bilayer patches have a typical width no more than one SiC terrace\cite{Jabakhanji2014} (see Methods). The hypothesis of geometric constraint only imposed by SiC steps could explain that the linear relationship between $R_\mathrm{H}$ and $R_\mathrm{xx}$ is remarkably independent of the magnetic field value.
\begin{figure}[h]
\begin{center}
\includegraphics[width=3.3in]{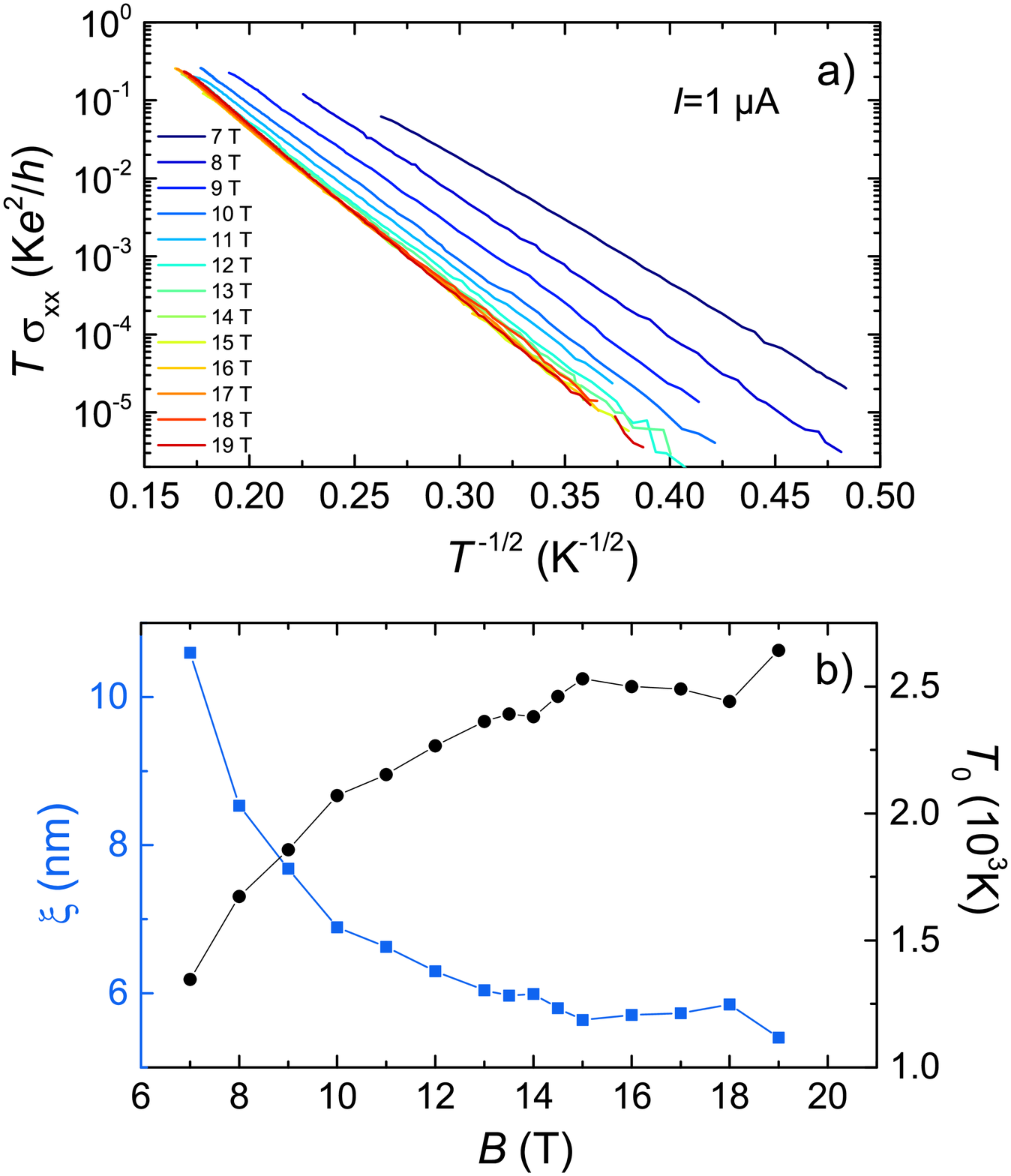}
\caption{\textbf{Analysis of the dissipation based on the VRH mechanism}. a) $T\sigma_\mathrm{xx}$ as a function of $T^{-1/2}$ in a semi-log scale for magnetic fields from 7 T to 19 T and in a temperature range from 4 K to 40 K. b) Temperature parameter $T_0$ (black circles, right axis) and localization length $\xi$ (blue squares, left axis) obtained from the adjustment of curves of figure a) by the VRH model as a function of the magnetic field. The dissipation in the QHE regime is well described by a VRH mechanism with a soft Coulomb gap. The continuous decrease of the localization length as the magnetic field increases, without showing a minimal value, explains the robustness of the Hall resistance plateau towards high magnetic fields.}\label{fig4}
\end{center}
\end{figure}

\subparagraph*{Dissipation through the variable range hopping mechanism.} To better understand the dissipation mechanism that alters the Hall quantization, Fig. \ref{fig4}-a) shows $\sigma_\mathrm{xx}(T)\times T$ plotted in logarithmic scale as a function of $T^{-1/2}$, where $\sigma_\mathrm{xx}=R_\mathrm{xx}/(R_\mathrm{xx}^2+R_\mathrm{H}^2)$ is the longitudinal conductivity. The linearity of the curves over five orders of magnitude allows the description $\sigma_\mathrm{xx}(T)\times T=\sigma_0(B)\exp[-(T_0(B)/T)^{1/2}]$ where $T_0(B)$ and $\sigma_0(B)$ are \emph{B}-dependent fitting parameters, as expected from a dissipation mechanism based on variable range hopping with soft Coulomb gap\cite{Shklovskii1984} that has  already been observed in exfoliated\cite{Giesbergvrh2009,Bennaceur2012} and epitaxial graphene\cite{Janssen2011}. Thermal activation does not manifest itself in the investigated temperature range up to 40 K. Figure \ref{fig4}-b) shows a sublinear increase of $T_0(B)$ as a function of \emph{B} with a saturation around 2500 K at the highest magnetic field. If we assume that $k_\mathrm{B} T_0(B)=Ce^2/(4\pi\epsilon_r\epsilon_0\xi(B))$ where $C=6.2$\cite{Lien1984,Furlan1998}, $k_\mathrm{B}$ is the Boltzmann constant, $\epsilon_0$ is the permittivity of free space, $\epsilon_r$ is the mean relative permittivity of the graphene on SiC covered by the P(MMA-MAA) copolymer ($\epsilon_r=(\epsilon_\mathrm{SiC}+\epsilon_\mathrm{P(MMA-MAA)})/2=7.25$, with $\epsilon_\mathrm{SiC}=10$\cite{Patrick1970} and $\epsilon_\mathrm{P(MMA-MAA)}=4.5$ is the value chosen, usually attributed to PMMA alone), then it is possible to determine the localization length $\xi(B)$ as a function of the magnetic field \emph{B}. Figure \ref{fig4}-b) shows that $\xi(B)$ continuously decreases from $\approx 10.5$ nm to $\approx 5.5$ nm between \emph{B}=7 T and \emph{B}=19 T but does not show a minimal value. This continuous decrease of $\xi(B)$ explains the robustness of the Hall resistance plateau towards high magnetic fields.

\begin{figure}[h!]
\begin{center}
\includegraphics[width=3.3in]{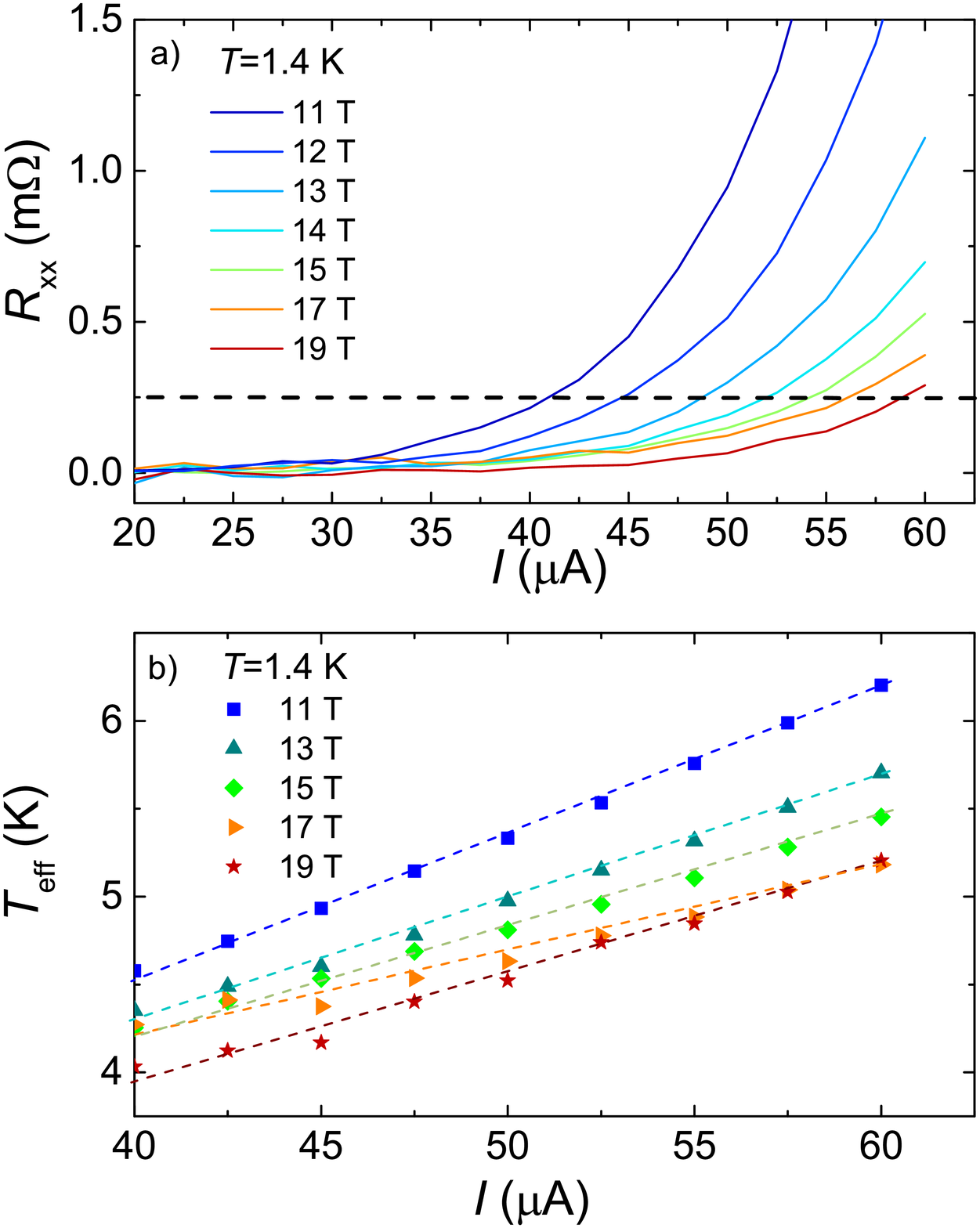}
\caption{\textbf{Current effect on dissipation}. a) $R_\mathrm{xx}$ as a function of the current at 1.4 K for different magnetic fields. The horizontal black dashed line indicates $R_\mathrm{xx}=0.25~\mathrm{m}\Omega$. b) Effective temperature $T_\mathrm{eff}$ as a function of \emph{I} (giving $\sigma_\mathrm{xx}(T_\mathrm{eff})=\sigma_\mathrm{xx}(I)$) for several magnetic fields \emph{B}. The linear relationship between $T_\mathrm{eff}$ and $I$ shows that the VRH mechanism with soft Coulomb gap explains the dissipation increase caused by the current.}\label{fig5}
\end{center}
\end{figure}

It is also interesting to know whether the VRH mechanism can explain the dependence of $R_\mathrm{xx}$ on the current reported in Figs. \ref{fig2} and \ref{fig5}-a). The VRH backscattering mechanism predicts that the current $I$ manifests itself as an effective temperature $T_\mathrm{eff}(I)=eR_\mathrm{H}I\xi/(2k_\mathrm{B}W)$ \cite{Furlan1998} where \emph{W} is the sample width in the hypothesis of a homogenous electric field. It results that $\sigma_\mathrm{xx}\propto \exp[-(I_0/I)^{1/2}]$ at \emph{T}=0 K where $I_0=2k_\mathrm{B}T_0 W/(eR_\mathrm{H}\xi)$ is a \emph{B}-dependent current parameter. For several $B$ values, the effective temperature $T_\mathrm{eff}(I)$ is determined by matching $\sigma_\mathrm{xx}(T_\mathrm{eff})=\sigma_\mathrm{xx}(I)$, where $\sigma_{xx}(I)$ is the conductivity measured as a function of the current and $\sigma_\mathrm{xx}(T)=(\sigma_0/T)\exp[-(T_0/T)^{1/2}]$ was determined previously from the data of Fig. \ref{fig4}-a). For all $B$ values, Fig. \ref{fig5}-b) shows a linear relationship between $T_\mathrm{eff}$ and $I$ as expected for the VRH mechanism. Moreover, the values of $T_\mathrm{eff}$ extracted in the investigated current range, belongs to a range of low temperatures ($\mathrm{< 7~K}$) where the VRH was demonstrated to explain the behavior of the longitudinal conductivity. From the slope of the $T_\mathrm{eff}(I)$ curves and the previous determination of $\xi(B)$, we can therefore extract an effective width $W_\mathrm{eff}\approx 7.5~\mathrm{\mu m}$, quite independent of the magnetic field, which is much smaller than the Hall bar channel width $W=100~\mu m$. This indicates an inhomogeneity of the current flow in the sample which holds up to large current values. In GaAs-QHRS supplied with high currents, several experiments based on the measurement of a linear dependence of the breakdown current of the QHE as a function of the Hall bar width, have strongly supported a homogeneous distribution of the current\cite{Jeckelmann2001}. On the other hand, sub-linear behaviors were also observed, generally in higher carrier-mobility samples\cite{Balaban1993,Meziani2004}. It turns out that the current distribution remains difficult to model because it is dependent on the microscopic details of the two-dimensional electron gas, notably of the length scale of inhomogeneities\cite{Jeckelmann2001,Furlan1998}. In exfoliated graphene on $\mathrm{SiO_2/Si}$, it was shown, for example, that large fluctuations of the carrier density caused by the presence of charged impurities close in the substrate lead to a drastic reduction of the breakdown current of the QHE\cite{Guignard2012}. In our graphene-based QHRS, Hall resistance measurements, performed at different places in the Hall bar do not reveal strong large-scale fluctuations of the carrier density (less than $10\%$). On the other hand, intermittent small bilayer patches existing along SiC edge steps constitutes inhomogeneities that could constraint the flowing of the current across constrictions and favor the existence of large local electric fields, resulting in a reduced effective width $W_\mathrm{eff}$. Fortunately, being of small size compared to the sample width, these bilayer patches are not able to short-circuit the edge states, an extreme effect which has been modeled\cite{Schumann2012,Lofwander2013} and recently observed in epitaxial graphene grown by sublimation of SiC\cite{Chua2014}. The proof is the accuracy of the quantized Hall resistance, demonstrated with a $10^{-9}$ relative measurement uncertainty, in the two G-QHRS considered in this work.

The 2D color plot of Fig. \ref{fig2} gives a direct visualization of \emph{I}(\emph{B}) curves at constant longitudinal resistance values. They are sub-linear, as highlighted by the black line which gives the evolution of the threshold current $I_C$ (which can be used to define a breakdown current of the QHE) above which $R_\mathrm{xx} > \mathrm{0.25~m\Omega}$. $I_C(B)$ continuously increases from $\mathrm{40~\mu A}$ to $\mathrm{60~\mu A}$ for \emph{B} varying from 10.5 T to 19 T. This corresponds to breakdown current densities varying from 0.4 A/m to 0.6 A/m if we assume, for the calculation, the $\mathrm{100~\mu m}$ width of the channel in between voltage terminals used to measure $R_\mathrm{xx}$. These values are similar to those measured in GaAs-QHRS but well below the best values reported in graphene grown by Si sublimation from SiC\cite{Alexander-Webber2013}. Nevertheless, we cannot omit that the injection of the current by the narrower $I_1$ terminal of $\mathrm{20~\mu m}$ width only could lead to a large underestimation of the breakdown current density. Furthermore, if we consider the effective width $W_\mathrm{eff} = \mathrm{7.5~\mu m}$ and the $I_C(B)$ values determined, we calculate higher breakdown current densities of 5.5 A/m at 10 T, 6.7 A/m at 14 T, and 8 A/m at 19 T, in agreement with values expected in graphene. The sub-linear evolution of $I_C$ as a function of \emph{B} can also be explained by the VRH mechanism. Given that $\sigma_\mathrm{xx}\propto \exp[-(I_0/I)^{1/2}]$, we indeed expect a sharp increase of the conductivity for $I_C\sim I_0$ with $I_0\propto \xi^{-2}$ (at this critical current the tiny variation of $\sigma_0$ becomes negligible). Figure \ref{fig2} indeed shows that $\xi^{-2}(B)$ (red squares) well adjusts to the $I_C(B)$ (black line).

\subsection*{Discussion}

In graphene, the combination of a large energy gap between LLs, the existence of a LL at zero energy and a moderate carrier mobility, which ensures a large mobility gap, are favorable to a wide extension of the $R_\mathrm{K}/2$ Hall resistance plateau, well beyond the magnetic field corresponding to $\nu_{n_0}=2$. Such wide and asymmetric (with respect to the magnetic field giving $\nu_{n_0}=2$) $R_\mathrm{K}/2$ Hall resistance plateaus have even been reported in some works either in exfoliated graphene\cite{Poumirol2010} or in epitaxial graphene grown on the C-terminated face of SiC \cite{Jouault2012}. Their quantization properties were characterized by a minimum of the longitudinal resistance occurring at a magnetic field corresponding to $\nu_{n_0}=2$. It results that an increase of the dissipation level, for example caused by an increase of the measurement current, tends to restore a symmetric shape of both the Hall and the longitudinal resistance with respect to the magnetic field giving $\nu_{n_0}=2$.

\begin{figure}[h!]
\begin{center}
\includegraphics[width=3.3in]{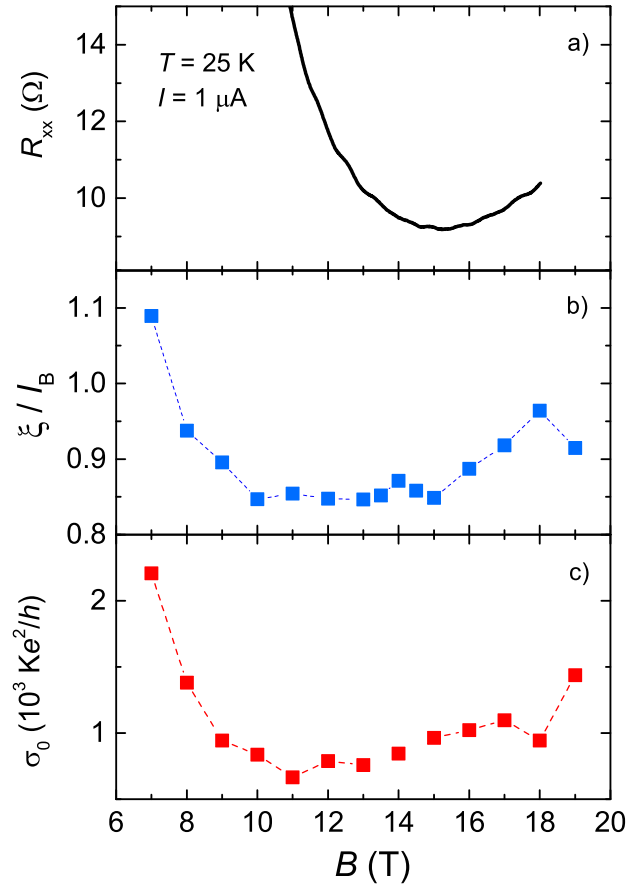}
\caption{\textbf{Correlation between quantization and localization}. a) $R_\mathrm{xx}$ \emph{versus} \emph{B} at \emph{T}=25 K and $I=1$ $\mu$A where we observe the presence of a tiny minimum at approximately 15 T. The ratio of the localization length to the magnetic length $\xi/l_B$ b) and the prefactor of the conductivity $\sigma_0$ c) extracted from the VRH analysis as a function of $B$. The minimum of $R_\mathrm{xx}$ occurs at the highest magnetic field where both $\xi/l_B$ and $\sigma_0$ have the lowest values. $\xi(B)$ is locked to the magnetic length $l_{B}$ within $10\%$ over the magnetic field range, from 10 T to 19 T, where the Hall resistance is accurately quantized with a $10^{-9}$ relative standard uncertainty.}\label{fig6}
\end{center}
\end{figure}

In the sample considered in this work, the magnetic field extension of the $R_\mathrm{K}/2$ Hall resistance plateau corresponds to a range of LL filling factor from $\nu_{n_0}=2.65$ (\emph{B}=5 T) down to $\nu_{n_0}=0.70$ (\emph{B}=19 T) if a carrier density $n_0$ constant with magnetic field is assumed. Moreover, the $\nu_{n_0}=2$ and $\nu_{n_0}=1$ LL filling factors should occur at \emph{B}=6.6 T and \emph{B}=13.2 T respectively. Measurements of $R_\mathrm{xx}$ at a low current value ($\mathrm{1~\mu A}$) as a function of \emph{B}, reported in Fig. \ref{fig6}-a) reveals the existence of a tiny minimum which occurs at $B\approx15~\mathrm{T}$ independently of the temperature between 1.3 K and 40 K, but not at \emph{B}=6.6 T, as would be expected in the hypothesis of a constant carrier density. On the other hand, $\nu=2$ at \emph{B}=15 T would mean a carrier density reaching $\mathrm{7.3\times10^{11}cm^{-2}}$ instead of $\mathrm{n_0=3.2\times10^{11}cm^{-2}}$. Moreover, this minimum is no more observable at larger currents of some tens of $\mu$A (see Fig. \ref{fig1}-c) and \ref{fig2}) while the increase of the dissipation should, in principle, reinforce its existence. On the contrary, we observe an exceptionally wide Hall resistance plateau, which remains accurately quantized with regards to the $10^{-9}$ relative standard measurement uncertainty over a 9 T magnetic field range, for macroscopic currents of several tens of $\mathrm{\mu A}$. This behavior is not in agreement with observations reported in previously discussed works ref.\cite{Poumirol2010,Jouault2012}. On the other hand, it is rather similar to what was observed in epitaxial graphene grown on the Si-terminated face of SiC by Tzalenchuk and co-workers\cite{Tzalenchuk2011}: an asymmetric Hall resistance plateau extending towards large magnetic fields that stays robust, and quantized to $R_\mathrm{K}/2$ within a relative uncertainty of a few $10^{-9}$ over 2.5 T, at large measurement currents. It was notably characterized by a continuous increase of the breakdown current of the QHE well beyond the magnetic field corresponding to $\nu_{n_0}=2$, as it is also observed in our sample (see Fig. \ref{fig5}-a)). This was explained by a pinning of the LL filling factor at $\nu=2$ caused by a charge transfer from the zero layer graphene (ZLG), specific to the growth on the Si-face of SiC, existing at the interface between the graphene and the substrate\cite{Kopylov2010}.

 To deepen our understanding, $\xi(B)$, reported in Fig. \ref{fig4}-b), was normalized by the magnetic length $l_B(B)=\sqrt{\hbar/eB}$ which describes the wavefunction characteristic size in the QHE regime. Fig. \ref{fig6}-b) shows that $\xi(B)/l_B(B)$ goes down when increasing \emph{B} up to 10 T, stays almost constant at a minimal value close to one between \emph{B}=10 T and \emph{B}=15 T, and then slowly increases at higher magnetic fields. Let us remark that the determination of $\xi(B)$ values slightly lower than the magnetic length, which was not expected, can be related to the assumptions regarding the values of the dielectric constants of SiC (the presence of the ZLG on SiC is not taken into account) and of P(MMA-MAA) ($\epsilon_\mathrm{P(MMA-MAA)}$ could be slightly different from $\epsilon_\mathrm{PMMA}$), as well as of the $C$ proportionality factor in the VRH expression of the conductivity (for instance, a larger value $C=7.4$ leads to $\xi(B)\geq l_B$ and might result from the partial inhomogeneity of our two-dimensional system caused by the presence of bilayer patches). In the magnetic field range from 10 T to 19 T, where the Hall resistance is accurately quantized with a $10^{-9}$ relative standard uncertainty, $\xi(B)$ remarkably stays very close, within $10\%$, to $l_B(B)$. Figure \ref{fig6}-c) also shows that the dependencies of $\sigma_0(B)$ and $\xi(B)/l_B(B)$ on \emph{B} are similar. It appears that the minimum of $R_\mathrm{xx}$, observed in Fig. \ref{fig6}-a) at $B\approx15~\mathrm{T}$, occurs at the highest magnetic field for which both $\sigma_0(B)$ and $\xi(B)/l_B(B)$ have the lowest values.

In the QHE regime, the localization length $\xi$ is expected to vary according to $\xi\propto \xi_0/\nu^{\gamma}$ (with $\gamma\approx 2.3$ \cite{Yoshioka1998} and $\xi_0$ a length depending on the disorder potential) for $\nu\leq 2$ and approaching $\nu=0$. In samples made of exfoliated graphene, this law was observed\cite{Bennaceur2012} to hold for $\nu$ values as high as 1.5. A lower bound value of $\xi_0$ is $l_B$, as predicted in case of short range disorder\cite{Polyakovprl1993}). We therefore expect $\xi(B)/l_B(B)$ higher than $1/\nu(B)^{\gamma}$ which increases for decreasing $\nu$ values and then diverges at $\nu=0$. For \emph{B} varying from 10 T to 15 T, although $\nu_{n_0}(B)$ decreases from 1.3 down to 0.9, $\xi(B)/l_B(B)$ is observed to stay constant. This is a first indication that $\nu(B)$ might stay close to $\nu=2$. Away from the LL center near integer filling factors, it was proposed that $\xi(B)$ should approach the classical cyclotron radius\cite{Fogler1998}. A localization length approaching $r_c=\hbar k_F/eB$ at integer LL filling factor, where $k_F$ is the Fermi momentum, was indeed observed in GaAs-based 2DEG\cite{Furlan1998}. In graphene, $r_c(B)$ can be written $r_c(B)=l_B(B)\sqrt{\nu/2}$ (since $k_F=\sqrt{\pi n_S}$). $r_c(B)$ is therefore proportional to $l_B(B)$ if $\nu(B)$ is constant (remarkably, one finds $l_B(B)$ for $\nu=2$).  The observation of $\xi(B)\sim l_B(B)$ therefore constitutes another argument suggesting that $\nu(B)$ could be pinned at $\nu=2$ from 10 T to 15 T, and then decreases slowly up towards 19 T.

 As discussed in Methods, structural characterization by low energy electron diffraction shows the existence of a ($6R\sqrt{3}\times 6R\sqrt{3}-R30^{\circ}$) reconstructed carbon-rich interface (ZLG) in our device\cite{Jabakhanji2014}. Thus, a transfer of charges from the ZLG leading to a pinning of $\nu(B)$ at $\nu=2$ is possible and could explain the large width of the observed Hall resistance plateau, the absence of minima for both the localization length and the longitudinal conductivity at \emph{B}=6.6 T. Using equations in ref.\cite{Tzalenchuk2011,Kopylov2010} derived from the balance equation $\gamma[A-(e^2d/\epsilon_\mathrm{0})(n_\mathrm{s}+n_\mathrm{g})-\epsilon_\mathrm{F}]=n_\mathrm{s}+n_\mathrm{g}$ describing the charge transfer, it is possible to reproduce a pinning at $\nu=2$ from \emph{B}=5.3 T up to \emph{B}=15.1 T with a zero magnetic field carrier density of $\mathrm{3.2\times 10^{11}cm^{-2}}$, considering \emph{A}=0.4 \emph{e}V, \emph{d}=0.3 nm, $\gamma=8.56\times 10^{12}~\mathrm{cm^{-2}(eV)^{-1}}$ and $n_\mathrm{g}=1.6\times10^{12}~\mathrm{cm^{-2}}$, where \emph{A} is the difference between the work functions of undoped graphene and ZLG, \emph{d} is the distance of the graphene layer to the ZLG, $\gamma$ the density of donor states in ZLG, and $n_\mathrm{g}$ the density of carriers transferred to the electrochemical gate. This is rather consistent with the experimental observations except that the analysis of the dependence of $\xi/l_B$ on \emph{B} rather indicates that the pinning of the filling factor should be effective at a higher magnetic field (\emph{B}=10 T). Further experimental and theoretical works are needed to better understand the peculiarities of the charge transfer in graphene grown by propane/hydrogen CVD on SiC. Thereupon, the reduced effective width $W_\mathrm{eff}$ over which the Hall potential drops, as determined from the analysis of the dissipation, could be an indication of some degree of inhomogeneity of the charge transfer.

\subparagraph*{Conclusion.} To summarize, we report on the Hall resistance quantization of the $\nu = 2$ plateau in a sample made of graphene grown by propane/hydrogen CVD on SiC. The agreement with $R_\mathrm{K}/2$ of the quantized Hall resistance, measured with a $10^{-9}$ relative standard uncertainty (1 s.d.) at \emph{T}=1.4 K, is demonstrated over a 9 T-wide magnetic field range extending from 10 T to 19 T. Moreover, the relative discrepancy between the quantized Hall resistances in the graphene sample and in a reference GaAs one is equal to $(-2\pm4)\times10^{-10}$. This constitutes a new proof of the universality of the QHE. The QHE physics of the wide quantized Hall resistance plateau is investigated using accurate specialized measurement techniques based on SQUID technology. From the characterization of the low dissipation, which is dominated by VRH, we determine that the localization length of states at Fermi energy stays locked to the magnetic length in the wide range of magnetic field where the Hall resistance is perfectly quantized. This can be explained by the pinning of the LL filling factor at $\nu=2$ caused by a charge transfer from the buffer layer (ZLG) at the interface between the graphene and the SiC. The analysis of the dissipation caused by the current reveals that the Hall electric field in the QHE regime is inhomogeneous across the sample, which could be linked to the structure of graphene grown by propane/hydrogen CVD on SiC . A second G-QHRS from a different graphene growth, measured at \emph{T}=1.3 K, is demonstrated to be $10^{-9}$-accurate at \emph{B}=10 T and over a magnetic field range wider than in usual GaAs-QHRS. This argues for the reproducibility of the fabrication method of G-QHRS which are able to substitute their GaAs counterparts, under the magnetic fields and low temperatures available in most national metrology institutes. This constitutes an essential step towards low magnetic-field G-QHRS setting the basis of low-cost and transportable G-QHRS in the near future. Given that the propane/hydrogen CVD on SiC is a scalable growth technique that produces high quality graphene meeting the demanding requirements of the resistance metrology, it is likely that it will be suitable for other electronic applications of graphene as well.
\subsection*{Methods}
\begin{figure}[h!]
\begin{center}
\includegraphics[width=2.5in]{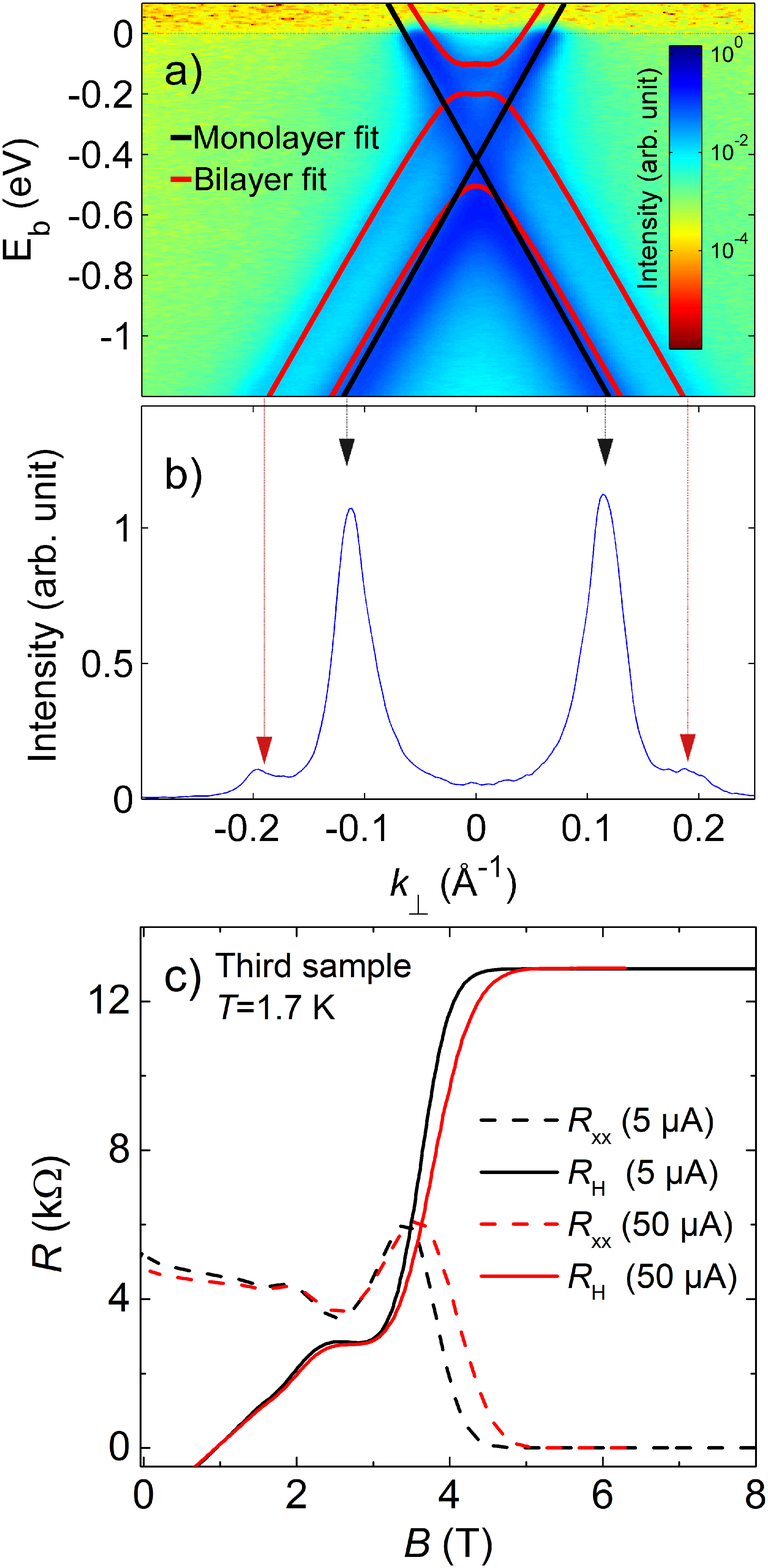}
\caption{\textbf{Complementary structural and electronic characterization}. a) Color-scale map of the ARPES intensity of the sample after outgassing at $\mathrm{500^{\circ}C}$. The intensity is plotted as a function of binding energy $E_b$ and momentum $k_\perp$ taken along the direction perpendicular to the $\Gamma\bar{K}$ direction in reciprocal space. The momentum reference is at the $\bar{K}$ point. The photon energy was 36 eV. The light was p polarized. The black and red solid lines are fits for monolayer and bilayer graphene respectively. b) ARPES intensity taken at $E_b=$ -1.2 eV, along  $k_\perp$, evidences the small bilayer contribution at $k_\perp=\pm 0.19$ \AA$^{-1}$. These ARPES measurements show that a graphene monolayer covers the whole SiC surface but approximately $10\%$ is covered by a second graphene layer. c) Longitudinal ($R_\mathrm{xx}$, in dashed lines) and transversal ($R_\mathrm{H}$, in solid lines) resistances as a function of \emph{B} for another Hall bar sample (400 $\mu$m by 1200 $\mu$m) fabricated from another piece ($5\times5~\mathrm{mm^2}$ size) of the same graphene wafer as first sample (same graphene growth run). The carriers density of the sample is $n_s=3.3\times10^{11}$ cm$^{-2}$ and the carrier mobility is $\mu=3300$ cm$^{2}$V$^{-1}$s$^{-1}$. The observation of similar QHE and the measurement of similar electronic properties in the first and third samples demonstrated the large scale homogeneity of the graphene growth and the repeatability of the fabrication process of samples having a few $\mathrm{10^{11}cm^{-2}}$ n-doping.}\label{fig7}
\end{center}
\end{figure}
\subparagraph*{Graphene growth.}
Graphene was grown by propane/hydrogen chemical vapour deposition\cite{Michon2010,MichonJAP2013} on the Si-face of a semi-insulating 0.16$^{\circ}$ off-axis 6H-SiC substrate from TanKeBlue. We used a horizontal hot-wall CVD reactor similar to that widespread in SiC electronic industry. A hydrogen/argon mixture ($23~\%$ of hydrogen)\cite{Michon2013} at a pressure of 800 mbar was used as the carrier gas during the whole process. The graphene growth was obtained by adding a propane flow ($0.04~\%$) for 5 min at a growth temperature  of 1550$^{\circ}$C. Before the lithography, the graphene was extensively analyzed (sample HT-MLG in ref. \cite{Jabakhanji2014}). Briefly, SiC steps of width 200 nm and height 0.75 nm were evidenced by AFM \cite{Jabakhanji2014}. Angle-resolved photoemission spectroscopy (ARPES) shows that a graphene monolayer covers the whole SiC surface but $\approx 10~\%$ is covered by a second graphene layer, Figs. \ref{fig7} a) and b). It grows discontinuosly and it is located mainly along SiC edge steps, forming  small bilayer patches of no more than 300 nm size. ARPES spectra also evidences high n-doping (10$^{13}$cm$^{-2}$) of the graphene monolayer, whose origin can be linked to the presence of a ($6R\sqrt{3}\times 6R\sqrt{3}-R30^{\circ}$) reconstructed carbon-rich interface detected by low-energy electron diffraction (LEED).

Finally, a remarkable homogeneity of the graphene film was evidenced in \cite{Jabakhanji2014} by the perfect superimposition of Raman spectra collected at different places of the sample. The lorentzian 2D peak and the normalized intensity of the G peak are typical of monolayer graphene. A notable D peak is observable but a large part originates for the underlying buffer layer \cite{Jabakhanji2014}. The homogeneity of the graphene film is confirmed by the measurement of very similar electronic properties (carrier mobility, similar QHE) in the main Hall bar studied and another (third sample considered in this work) fabricated from a different piece ($5\times5~\mathrm{mm^2}$ size) of the same graphene wafer (see Fig. \ref{fig7}-c)). Moreover, the structural properties of the graphene were demonstrated to be repeatable and well-controlled by the growth parameters (pressure, temperature, propane and hydrogen flow)\cite{Michon2010,Michon2013,MichonJAP2013}. This is evidenced by the measurement of a second $10^{-9}$-accurate G-QHRS from \emph{B}=10 T fabricated from a different graphene growth (several months later), as mentioned in subsection Resistance quantization.

\begin{figure}[h!]
\begin{center}
\includegraphics[width=2.5in]{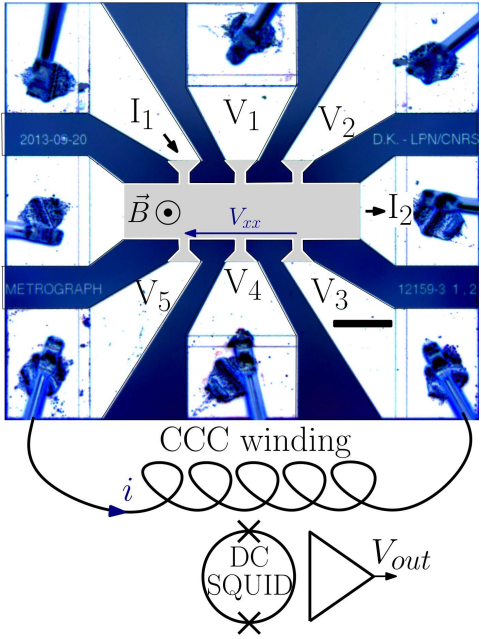}
\caption{\textbf{Schema of the $R_\mathrm{xx}$ measurements using a CCC}. The sample is biased with a DC current \emph{I} and a 2065 turns winding of a CCC is connected to the two voltage terminals. The voltage drop $V_\mathrm{xx}$ gives rise to the circulation of a current \emph{i} in the winding. The longitudinal resistance $R_\mathrm{xx}$ is given by $R_\mathrm{xx}=(i/I)R_\mathrm{H}$.}\label{fig8}
\end{center}
\end{figure}

\subparagraph*{Sample fabrication.} The graphene sample was annealed in vacuum (a few $10^{-4}$ hPa pressure) for 1 min at $\mathrm{500^{\circ}C}$ (ramp of 500 s). The sample was left to cool down to below $\mathrm{100^{\circ}C}$ in vacuum over a few minutes. Subsequently, it was covered with polymethylmethacrylate (PMMA) for protection. The Hall bars were patterned using electron-beam lithography with PMMA resist and oxygen reactive ion etching (RIE). Ohmic contacts to the graphene layer were formed by depositing a Pd/Au-(60/20 nm) bilayer in an electron beam deposition system, using an ultrathin Ti layer for adhesion. Thicker Ti/Au-(20/200 nm) bonding pads were formed in a subsequent step, where a RIE etch was performed prior to metal deposition for better adhesion of the metal pads to the SiC substrate. The Hall bar has a width of $\mathrm{100~\mu m}$ and a total length of $\mathrm{420~\mu m}$. It has three pairs of Hall probes, separated by $\mathrm{100~\mu m}$ (see inset Fig. 1b and Fig. \ref{fig8}). Finally the sample was covered for protection by 300 nm of poly(methylmethacrylate-\emph{co}-methacrylate acid) copolymer (MMA (8.5) MAA EL10 from  Microchem) and 300 nm of poly(methylstyrene-\emph{co}-chloromethylacrylate) (ZEP520A from Zeon Chemicals) resist. The ZEP520A resist is known to reduce the electron density under UV illumination\cite{Tzalenchuk2011,Lara-Avila2011}. Nonetheless, no illumination was done in our case. Although not fully understood, the process leading to low carrier density is reproducible. Fig. \ref{fig7}-c) which reports the QHE in another Hall bar sample (third sample considered in this work) characterized by values of carrier mobility and density quite close to those of the main sample studied, illustrates the repeatability of the fabrication process of samples having a few $\mathrm{10^{11}cm^{-2}}$ n-doping.

\subparagraph*{Measurement techniques.} The Hall resistance $R_\mathrm{H}$ of a QHRS is compared to the $100~\Omega$ resistance of a transfer wire resistor using a resistance bridge based on a cryogenic current comparator (CCC). The CCC is a perfect transformer which can measure a current ratio in terms of the winding number of turns ratio with a relative uncertainty as low as a few $10^{-11}$.
Its accuracy relies on a flux density conservation property of the superconductive toroidal shield (Meissner effect), in which superconducting windings are embedded. Owing to a flux detector based on a DC superconducting quantum interference device (SQUID), the current noise resolution of the CCC is $\mathrm{80~pA.turn.Hz^{-1/2}}$.

For measurements reported in Fig. \ref{fig1}-a), the resistance bridge operates in direct current (DC) mode (the current is reversed every 35 s) and is equipped with a EMN11 nanovoltmeter as a null detector. The QHR and the $100~\Omega$ resistor are connected in series with a 2065 turns winding and a 16 turns winding respectively. $R_\mathrm{xx}$ is determined using an EMN11 nanovoltmeter to detect the longitudinal voltage $V_\mathrm{xx}$ resulting from the circulation of a DC current in the Hall bar.

For measurements reported in Fig. \ref{fig2} and \ref{fig5}-a) and for two measurements reported in the text (Resistance quantization section) carried out at \emph{B}=10 T and \emph{B}=19 T, the sample is biased with a DC current \emph{I} and $R_\mathrm{xx}$ is measured using the CCC (see Fig. \ref{fig8}). A 2065 turns winding of the CCC is connected to the two voltage terminals. The longitudinal voltage $V_\mathrm{xx}$ gives rise to the circulation of a current \emph{i} in the winding of the CCC which is used as a current amplifier with a SQUID operating in internal feedback mode. The output of the SQUID electronics is measured with an Agilent 3458A multimeter. The current noise resolution is approximately $\mathrm{40~fA/Hz^{1/2}}$ which results in a voltage noise resolution of $\approx\mathrm{0.5~nV/Hz^{1/2}}$. The longitudinal resistance $R_\mathrm{xx}$ is then given by $R_\mathrm{xx}=(i/I)R_\mathrm{H}$ since the two-terminal impedance seen by the winding is very close to $R_\mathrm{H}$ on the $\nu=2$ plateau.

For data reported in Fig. \ref{fig3} and \ref{fig4}, quick and accurate measurements are carried out while the temperature of the sample is swept from 40 K down to 3 K. The Hall bar is then supplied with an AC (2 Hz frequency) current $I=1~\mathrm{\mu A}$, controlled by the reference voltage of a Signal Recovery 7265 lock-in detector. The Hall resistance $R_\mathrm{H}$ is measured using the resistance bridge replacing the EMN11 nanovoltmeter used in DC to measure the voltage balance by a Celians EPC1 AC low-noise amplifier whose output is connected to the lock-in detector. $R_\mathrm{xx}$ is measured using the CCC as in Fig. \ref{fig2} and \ref{fig5}-a), except that the output of the SQUID is connected to the lock-in detector.

\subparagraph*{References}
\providecommand{\noopsort}[1]{}\providecommand{\singleletter}[1]{#1}%

\subparagraph*{Acknowledgments}
We wish to acknowledge D. Leprat for technical support, D. Mailly for advices about nano-fabrication, M. Paillet, A. Zahab, A. Tiberj, J.-R. Huntzinger, W. Desrat for advices and fruitful discussions, F. Bertran, P. Le F\`evre and A. Taleb-Ibrahimi for their support at the SOLEIL synchrotron radiation facility. This research has received funding from the Agence national de la Recherche (ANR), Metrograph project (Grant No. ANR-2011-NANO-004). It has been partly performed within the EMRP (European Metrology Research Program), project SIB51, Graphohm. The EMRP is jointly funded by the EMRP participating countries within EURAMET (European association of national metrology institutes) and the European Union. The SOLEIL synchrotron radiation facility is acknowledged for providing beamtime under project n$^\circ$20120817.\\

\subparagraph*{Author contributions}
W. P and F. S. planned the experiments. A. M. fabricated the graphene layer. A. M., T. C, M. Z and M. P developed the growth technology. D. K. fabricated the Hall bars. F. L., R. R-P., F. S., W. P. conducted the electrical measurements. B. J. and A. M. performed ARPES measurements. D. K., B. J., C. C and O. C carried out complementary electrical measurements. F. L., R. R-P., F. S., W. P. analyzed the data.  W. P., F. S, F. L, R. R-P, B. J, A. M, D. K wrote the paper with all authors contributing to the final version.

\subparagraph*{Competing financial interests}
The authors declare no competing financial interests.

\subparagraph*{Corresponding author}
Correspondence to: Wilfrid Poirier (wilfrid.poirier@lne.fr)

\end{document}